\documentclass[10pt,preprint]{aastex}

\usepackage{emulateapj5}
\usepackage{amsmath}

\begin{document}

\title{Ejection of Supernova-Enriched Gas From Dwarf Disk Galaxies}

\author{P. Chris Fragile, Stephen D. Murray}
\affil{University of California,
Lawrence Livermore National Laboratory, P.O. Box 808, Livermore, CA 94550}

\and

\author{Douglas N. C. Lin}
\affil{University of California,
Lick Observatory, Santa Cruz, CA  95064}


\begin{abstract}
We examine the efficiency with which supernova-enriched gas may be
ejected from dwarf disk galaxies, using a methodology previously
employed to study the self-enrichment efficiency of dwarf spheroidal
systems.  Unlike previous studies that focused on highly concentrated
starbursts, in the current work we consider discrete supernova events
spread throughout various fractions of the disk.  We model disk
systems having gas masses of 10$^8$ and 10$^9$~M$_\odot$ with
supernova rates of 30, 300, and 3000~Myr$^{-1}$.  The supernova events
are confined to the midplane of the disk, but distributed over radii
of 0, 30, and 80\% of the disk radius, consistent with 
expectations for Type II supernovae.  In agreement with earlier studies,
we find that the enriched material from supernovae is largely lost
when the supernovae are concentrated near the nucleus, as expected for
a starburst event.  In contrast, however, we find the loss of enriched material
to be much less efficient when the supernovae occur
over even a relatively small fraction of the disk.  The difference is
due to the ability of the system to relax following supernova events
that occur over more extended regions.  Larger physical separations
also reduce the likelihood of supernovae going off within low-density
``chimneys'' swept out by previous supernovae.  We also find that, for
the most distributed systems, significant metal loss is more likely to
be accompanied by significant mass loss. A comparison with theoretical
predications indicates that, when undergoing self-regulated star
formation, galaxies in the mass range considered shall
efficiently retain the products of Type~II supernovae.
\end{abstract}


\section{Introduction}
\label{sec:intro}

In cosmological models dominated by cold dark matter (CDM), the
amplitude of fluctuations increases toward shorter wavelengths.  Small
galaxies are therefore the first to form, and large galaxies
subsequently form through the merger of smaller systems
\citep[e.g.][]{WR78,BFPR84,C94,KNP97,NFW97}.  The very first
generations of star formation, and hence the initial enrichment of the
interstellar and intergalactic medium, therefore likely occurred
within dwarf galaxies.  Much recent attention has been paid to
understanding the gas evolution and first generations of stars within
such systems \citep[e.g.][]{DS86, Katzetal96, QKE96, Tegetal97, WHK97,
MF99, SGV99, SGP03, FMAL03}.

Observational work is also shedding light upon the early evolution of
dwarf galaxies.  Because the comoving density of damped Ly$\alpha$
systems (DLAs) at $z>$2 is comparable to that of all the ordinary
matter in galactic disks today, it is tempting to identify the DLAs as
the objects that evolved into the present-day L$_\ast$ galaxies
\citep{Kauf96}.  The observationally inferred star-formation rates
per unit surface area in such systems are similar to that in the Milky
Way today \citep{WPG03}.  Recent observations find that the DLA
metallicity evolves significantly with redshift.  The overall
metallicities are, however, well below that expected from
observationally inferred star-formation rates, implying a significant
loss rate of enriched gas from these systems \citep{PGWCD03}.

In addition to governing the early enrichment of the universe, star
formation within dwarf systems affects their dynamical evolution as
they merge to form more massive galaxies.  One important issue is the
``angular momentum problem.''  In CDM models which include only
cooling, the first objects to form rapidly cool and contract.
Numerical simulations show a tendency for dwarf galaxy building blocks
to undergo dynamical friction and to lose a large fraction of their
initial angular momentum prior to merging, leading to the formation of
massive galaxies that are much more compact and have much less angular
momentum than observed \citep{SGV99}.  Some cosmological models have
avoided this problem by invoking {\it ad hoc} star-formation rates
within dwarf systems, in order to heat their gas and maintain large
radii.

In earlier work, we have explored star formation that is
self-consistently governed by negative feedback, in a self-regulating
process.  That picture has had good success in matching the
star-formation rates seen in spheroids and disks \citep{LM92,LM00},
and in matching observational data for dwarf spheroidal systems
(dSphs) \citep{DLM03}.

For dSphs, \cite{DLM03} find that star formation may occur in bursts.  In
such low mass systems, even a small burst of supernovae can lead to a 
substantial loss of enriched gas \citep{FMAL03}.  In more massive
dwarf disk galaxies, however, where the dynamical times exceed the
characteristic timescales for star formation and feedback effects, star
formation is expected, in the absence of dynamical interactions between
galaxies, to be a much more steady-state process.

The ejection of enriched gas from dwarf disk systems has been examined
in many previous studies \citep{DG90,DH94,ST98,MF99,SGP03}.  In those
studies, it was found that the efficiency with which enriched material
is ejected from dwarf disk galaxies may be quite high.  Even in
systems with gas masses as high as 10$^9$~M$_\odot$
(10$^{10}$~M$_\odot$ total mass), the ejection efficiency of enriched
material was 97\% for a supernova rate of one per 30,000~yr
\citep{MF99}.  This efficiency is especially striking given that,
measured per unit gas mass, the supernova rates assumed are more than
two orders of magnitude below that of the Milky Way today.  The high
ejection efficiency in the modeled dwarf systems is the result of
both the shallower potential of the lower mass galaxy and the mode of
star formation.  In \citet{MF99}, the supernova energy was input as a
continuous luminosity source at the center of the model galaxies, and
so the systems were unable to relax between discrete supernova events.
\citet{SGP03} allowed star formation to occur with high local
efficiency whenever the gas density exceeded a critical value,
resulting in strong bursts of star formation that also led to
efficient ejection of enriched gas.  These models, therefore, best
represent the evolution of concentrated starbursts within the dwarf
disk systems and indicate that even mild starbursts may lead to
significant loss of enriched material.

In general, however, star formation is not limited to concentrated
regions within young galaxies.  Although it is expected to occur most
strongly within regions of higher gas surface density, it shall occur
throughout the disk, as is supported both by observational data
\citep{GH87,Kennicutt89} and self-regulating models \citep{LM92}.
Because self-regulating star formation is expected to dominate the
life history of galaxies, it is of interest to see how efficiently the
enriched material produced by this more quiescent process is retained.

We note here that our simulations restrict supernovae to occur in
the regions of our models where the gas density is greatest, namely
the midplanes.  As we discuss below, this is consistent with
expectations for Type II supernovae, since they occur mostly in
regions of high star formation efficiency.  Type I supernovae, on the
other hand, which are believed to be responsible for most of the iron
production within galaxies, come from older objects, which may have
migrated significantly from their birthplaces and so may occur at
heights well above the disk plane.  In a follow up paper, we will
consider the impact of Type I supernovae and the implication on the
contamination of clusters of galaxies.

In this paper, we extend our earlier work on the ejection of enriched
gas from dSphs \citep{FMAL03}, by examining three-dimensional
simulations of supernova energy input to dwarf disk galaxies, in which
the supernovae occur as discrete events, spread out over significant
fractions of the disk.  We describe our numerical method and the setup
of the models in \S~\ref{sec:method}.  The results of the models are
discussed in \S~\ref{sec:results}, and conclusions for the evolution
of dwarf disk systems are discussed in \S~\ref{sec:discussion}.

\section{Numerical Method}
\label{sec:method}

\subsection{Numerical Code}
\label{sec:methodcode}

The models discussed below have been computed using Cosmos, a
massively parallel, multidimensional, 
radiation-chemo-magneto-hydrodynamics code 
for both Newtonian and relativistic flows developed at Lawrence Livermore
National Laboratory.  Tests of the Newtonian hydrodynamics options and of the
internal physics relevant to the current work are presented in \cite{AFM03} 
and shall not be discussed in detail here.

Because we shall be examining the effects of having multiple supernova
explosions at random locations throughout the cores of dwarf galaxies, 
our models are run in
three-dimensions on a Cartesian mesh.  The simulations 
are run on a $256\times256\times128$ grid with a physical box size of 
30 kpc $\times$ 30 kpc $\times$ 15 kpc.  Each zone, therefore, measures 
117 pc on a side.  
We explore the impact of numerical resolution on our results by also 
performing one simulation at a resolution of $128\times128\times64$.  
We use flat (zero-gradient) boundary conditions at each 
of the outer boundaries and a reflective boundary along the midplane 
of the disk.
Because highly evacuated regions can be a problem for numerical 
hydrodynamics codes, we maintain a strict density floor at 
$10^{-5}$ of the initial midplane density of the galaxy.

The code parameters used for the present work are similar to those
used previously in studying the evolution of enriched material in dwarf
spheroidal galaxies \citep{FMAL03} and in jet-cloud interactions
\citep{FMAV04}.  Chemistry is not followed in these simulations, and cooling
is given by an equilibrium cooling function, assuming low metallicity 
($Z=0.03Z_\odot$).  In
order to simulate the effects of equilibrium heating, cooling is suppressed
below $10^4$~K.  Due to our inability to resolve the cooling regions behind
shocks, cooling is also restricted whenever $Q/P>0.1$, where $Q$ is the
scalar artificial viscosity and $P$ is the thermal gas pressure.  This
should help prevent overcooling in the unresolved post-shock gas, although
this was not found to have a strong effect upon the evolution of 
dwarf spheroidal galaxy models \citep{FMAL03}.

\subsection{Galaxy Model}
\label{sec:setup}

The galaxies are modeled as non-self gravitating gas within a fixed 
dark-matter (DM) potential.  The omission of self-gravity is reasonable,
given that the baryonic-to-dark matter ratio of the systems is $\sim0.1$.
In order to facilitate comparisons with previous work, we adopt the same
form for the galactic potential as used by both \citet{MF99} and \citet{ST01}.
The spheroidal DM mass distribution is given by
\begin{equation}
\rho_d (r) = \frac{\rho_c}{1 - (r/R_c)^2}~,
\end{equation}
where the central density ($\rho_c$) and scale radius ($R_c$) are 
\citep{ST01} 
\begin{eqnarray}
\rho_c & = & 6.3 \times 10^{10} \left(\frac{M_d}{M_\odot} \right)^{-1/3} h^{-1/3} ~M_\odot ~\mathrm{kpc}^{-3} \label{eq:rhoc} \\
R_c & = & 8.8 \times 10^{-6} \left(\frac{M_d}{M_\odot} \right)^{1/2} h^{1/2} ~\mathrm{kpc}~, \label{eq:r0}
\end{eqnarray}
where $M_d$ is the mass of the dark matter halo and 
$h$ is the Hubble constant in units of 100 km s$^{-1}$ Mpc$^{-1}$.  
We adopt $h=0.65$, consistent with measurements using Type Ia supernovae
\citep{GMS00} and gravitational lensing \citep{WS00}, and slightly smaller
than the value determined by recent Cepheid measurements
\citep{Fetal01}.
The tidal radius of the dark matter halo ($R_t$) is given by
\citep{MF99}:
\begin{equation}
R_t = 0.016 \left(\frac{M_d}{M_\odot} \right)^{1/3} h^{-2/3} ~\mathrm{kpc}~. \label{eq:rh}
\end{equation}
The total mass of the halo is
\begin{equation}
M_d = 4\pi \int_0^{R_t} \rho_d(r) r^2 dr = 4\pi\rho_c R_c^3 (x_t - \arctan x_t) ~,
\end{equation}
where $x_t=R_t/R_c$, 
is related to the mass of the visible (gas plus stellar) matter ($M_g$) 
by \citep{PSS96}
\begin{equation}
\frac{M_d}{M_g} \simeq 34.7 \left(\frac{M_g}{10^7 M_\odot}\right)^{-0.29}~.
\end{equation}
The gravitational potential of the halo is 
\begin{equation}
\phi (r) = 4\pi G \rho_c R_c^2 \left[ \frac{1}{2} \ln (1+x^2) + 
\frac{\arctan x}{x} \right] ~,
\end{equation}
where $x=r/R_c$.  The interstellar gas density is then 
distributed according to \citep{ST98}
\begin{eqnarray}
\rho & = & \rho_0 \exp \left[ \frac{3}{2} \left( \frac{V_{esc}}{c_s}\right)^2 
\chi \right] ~,
\label{eq:rho}
\end{eqnarray}
where $\rho_0$ is the gas density at the galactic center and
$V_{esc}=(2GM_d/R_t)^{1/2}$ is the escape velocity at the halo
boundary ($R_t$).  We adopted the equation of state of an ideal gas
with cooling and heating such that the effective sound speed evolves 
with time.  Here $c_s$ is the initial value of the sound speed.
The function $\chi$ determines the density 
distribution in the (vertical) direction normal to the disk.  
It has the form \citep{ST98}
\begin{equation}
\chi = F(r) - F(\varrho) ~,
\end{equation}
where $\varrho$ is the cylindrical radius and 
\begin{eqnarray}
F(r) = & & 1 + \frac{4\pi \rho_c R_c^3}{M_d} x_t \left\{ \frac{1}{2} 
[\ln(1+x_t^2)-\ln(1+x^2)] \right. \nonumber \\
 & & \left. + \frac{\arctan x_t}{x_t} - 
\frac{\arctan x}{x} \right\} ~.
\end{eqnarray}
The gas disk extends out to a cutoff radius, which depends upon the gas
mass following the relation \citep{FT00}
\begin{equation}
R_d = 3 \left( \frac{M_g}{10^7 M_\odot} \right)^{0.338} \mathrm{kpc}~.
\end{equation}
Vertically, the disk extends to a height determined by $P=P_{IGM}$,
where $P=\rho c_s^2$ is the interstellar gas pressure.  The value of
$\rho_0$ in equation \ref{eq:rho}, which sets the midplane density, 
is scaled to give the desired gas mass $M_g$
within $R_d$.  This density distribution is chosen primarily for the
purpose of comparison with other calculations. Although it does not
represent that of an exponential disk, the gas profile is a
reasonable approximation to dwarf galaxies.  In the midplane of the
galaxy, the gas is supported radially against gravity entirely by
rotation.  The circular velocity of the galaxy is
\begin{equation}
v_c^2 (r) = r \frac{\partial \phi}{\partial r} = 
\frac{4 \pi G \rho_c R_c^2}{x} (x - \arctan x) ~.
\end{equation}
Our models are, therefore, completely specified by choosing $M_g$,
$c_s$, and $P_{IGM}$.  In this work, all models use $c_s=10$ km
s$^{-1}$ and $P_{IGM}/k=1$ cm$^{-3}$ K. In Models 1-7, we choose $M_g =
10^9 M_\odot$, which corresponds to $M_d \sim 10^{10} M_\odot$, $v_c
\sim 35$km s$^{-1}$, $R_c \sim 1$ kpc, $R_d \sim 14$ kpc, and $R_t
\sim 30$ kpc.  The gas mass in Model 8 is an order of magnitude smaller.  
The masses of these models are smaller than
$L_\ast$ galaxies but much larger than dwarf spheroidals.

\subsection{Supernova Input}
\label{sec:supernovae}

In this work supernova events are simulated by adding either 10$^{51}$
or 10$^{52}$~erg of internal energy to the gas over finite regions; 
other fluid variables remain unchanged.
The internal energy is injected over an approximately spherical volume
with a ``top-hat'' profile of 1-4 zones radius.  Normally a radius of 
2 zones is used, yielding an initial supernova gas temperature of 
$T_{SN} \gtrsim 10^6$ K for a $10^{52}$ erg event.  
For the low metallicities used in these 
simulations, this lies close to the minimum of the cooling function, 
helping prevent the supernova gas from cooling too quickly.  
A supernova radius of 1 zone is used for our low-resolution 
($128\times128\times64$) simulation, in order to keep the size of 
the input region as 
similar as possible to the higher resolution equivalent.  We only 
use a supernova radius of 4 zones for the highest luminosity model 
considered (Model 7).  Although supernova events at the beginning 
of that simulation may cool too rapidly, the high event rate 
ensures that later supernovae are set off in previously evacuated regions 
and are not subject to excessive cooling.

Whenever supernova events are set off, a passive tracer is added to
the same zones to which the energy is initially added.  We use this
tracer to track the ejection of enriched material as described in 
\S \ref{sec:loss}.

Runs with supernova
events of 10$^{52}$ erg increase the chances of a single event
successfully expelling enriched material from the galaxy.  This also 
helps ensure that a reasonable minimum floor temperature is 
maintained for the supernovae ($T_{SN} \gtrsim 10^6$ K).  This treatment
simulates the fact that multiple supernovae are often expected
to occur in localized regions, centered upon OB associations.  The
effective local energy input in these regions is therefore expected to
exceed that of a single supernova.  However, the more energetic events
can be computationally expensive to evolve for a time-explicit code
such as Cosmos, especially when supernova events overlap, as occurs in
either our most confined runs or highest luminosity cases.  Therefore,
some runs input 10$^{51}$ erg per event.  For these runs most of the 
supernovae go off in previously evacuated regions so they have a 
relatively high initial temperature, although 
some of the early supernovae may be much cooler.

\begin{figure}
\epsscale{0.8}
\plotone{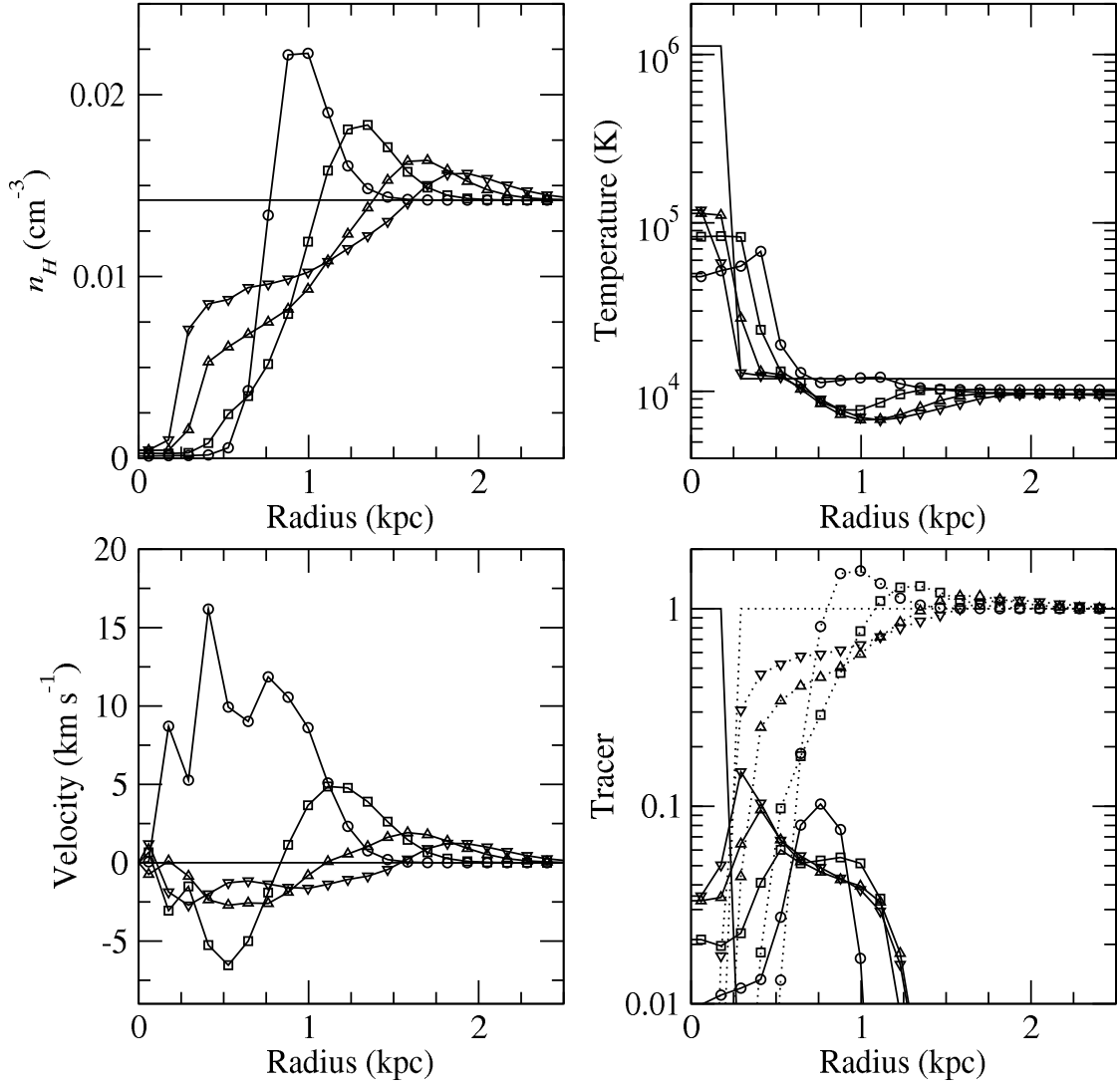}
\caption{Density, temperature, velocity, and tracer distribution 
of a single supernova event in a homogeneous background.  
The data are from $t=0$, 25 ({\it circles}), 50 ({\it squares}), 
75 ({\it triangles}), and 100 ({\it inverted triangles}) Myr.  
The tracer plot tracks both enriched material ({\it solid lines}) and 
background gas ({\it dotted lines}).
}
\label{fig:SNcomp}
\end{figure}

\begin{figure}
\epsscale{0.5}
\plotone{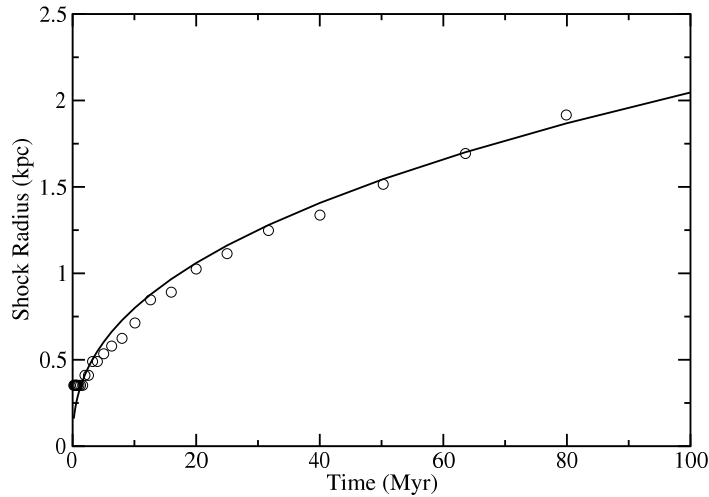}
\caption{Shock radius as a function of time for a single supernova 
event in a homogeneous background.  The data ({\it circles}) are 
best fit with a power-law of the form $r_{sh}\propto t^{0.41}$ 
({\it solid curve}), close to the Sedov-Taylor solution.
}
\label{fig:SNshock}
\end{figure}

To illustrate some of the properties of these supernova events, 
we first simulate a single event in a homogeneous background.  This 
simulation uses the same grid spacing as our standard models (117 pc).  
Gravity is not considered and the background is 
set to a uniform density and temperature equal to the initial midplane 
values of our $10^9 M_\odot$ galaxy.  We evolve the model 
for 100 Myr, the same as most of our galaxy models.  In Figure 
\ref{fig:SNcomp}, we plot several fluid variables as a function of 
radius at equally spaced intervals of time.  Here we comment on a few 
key observations from that simulation.  First, it is apparent that 
much of the 
cooling of the supernova gas is dominated by adiabatic expansion, 
although some radiative cooling does take place 
in the dense post-shock shell.  
Also, the supernova expansion closely follows the Sedov-Taylor solution.  
This is best illustrated in Figure \ref{fig:SNshock}, which shows 
the radius of the shock front ($r_{sh}$) as a function of time.  The data are 
best fit with a power-law of the form $r_{sh}\propto t^{0.41}$, 
very close to the Sedov-Taylor solution $r_{sh}\propto t^{2/5}$.  
One concern, however, is the noticeable mixing of tracer material in the 
post-shock shell (lower right panel of Figure \ref{fig:SNcomp}).  
Although, some mixing is expected due to 
hydrodynamic (Rayleigh-Taylor and Kelvin-Helmholtz) instabilities, 
there is also the possibility of anomalous mixing due to numerical 
diffusion.  Since the tracer is a surrogate for enriched gas mass in these 
simulations, this suggests that the metal ejection efficiency could 
be artificially reduced due to excessive mixing of cold, ``heavy'' 
background gas with the supernova gas.  This is one motivation for 
testing our results at a different resolution, since this will provide 
some indication of how problematic numerical diffusion is.  
We also point out that, by $t\gtrsim 10$ Myr or an expansion radius of 
$r_{sh}\gtrsim 800$ pc, the post-shock gas in the 
supernova is already traveling slower than the local escape speed 
at the core of our galaxy models.  This is certainly consistent with 
our findings that a single supernova event (even a $10^{52}$ erg event) 
is not sufficient to eject 
enriched material from the galaxy potentials studied here [nor 
was a single $10^{51}$ erg event sufficient for the much smaller 
dwarf spheroidal galaxies studied in \citet{FMAL03}].  Multiple 
overlapping events or pre-existing voids are crucial to propelling 
the enriched material out of the galaxy.

In the remaining simulations, 
the supernova events are distributed randomly in space and time,
subject to the following restrictions: the average rate of events is a
fixed parameter within each model; all supernovae are restricted to
the midplane of the galaxy; and supernovae are restricted to lie
within a specified radius of the galactic center, with their
distribution weighted by the enclosed area.  Confining supernova
events to the midplane makes it more reasonable to employ a reflective
boundary as we have done.  This also maximizes the probability that
the galaxy will be able to contain the energy of each event and retain
the enriched material.  Because star formation is expected to be
strongly concentrated where the gas density is greatest, limiting
supernova events to the midplane is also a reasonable approximation to
their actual distribution (at least for Type II supernovae).

\subsection{Model Parameters}
\label{sec:methodgalaxy}

The parameters of the models are listed in Table~\ref{tab:models}.
For each model, we list the gas mass of the galaxy ($M_g$), the energy
input rate, the energy per supernova event, the fractional radius of
the disk over which the supernovae occur ($R_{SN}/R_d$), the time
interval over which the supernovae are initialized, and the total
simulation time.  

The parameters for Model~1 are selected so as to closely match the
highest mass, highest luminosity model of \citet{MF99}.  The primary
difference is that the supernovae occur discretely in time, as is the
case in all our models.  In order to more closely match the results of
\citet{MF99}, who use a continuous input of energy, the supernova
events in Model 1 have 10$^{51}$ erg each.  The high supernova
frequency helps to ensure that the nuclear gas does not have
substantial time to relax between events, which should help to
minimize the differences between this model and the earlier work.  
The duration of supernovae input phase is set by two factors.  The
main sequence lifetime of the smallest stars that become supernovae is
50~Myr \citep{Strothers72,McCray87}.  This sets the minimum
timespan over which supernovae shall occur, in the limit of an
instantaneous burst of star formation.  The timespan shall be larger
if star formation occurs over an extended time period.  One
characteristic timescale is the dynamical time of the core, $\sim$30~Myr
for Model~1.  The total timescale over which supernovae may occur could
therefore be close to 100~Myr.  We adopt the minimum time scale for 
Model~1, and note that had we injected the same supernova energy over the
longer timescale, the loss of enriched gas might be reduced from that found
in Model~1.

Models~2-7 use the same unperturbed system as Model~1.  Model 2 is
exactly the same as Model 1 except that the supernovae are all set off
at a single point half-way out in the disk.  The scale height of the
gas increases with radius, and so this model examines the ability of a
non-nuclear starburst to expel enriched material, given that it must
work its way through a larger volume of interstellar gas.  Models~3
and 4 use the same energy input rate as Model 1 but with ten times
more energy per event.  In Model 3, the supernova events are spread
out over 80\% of the galactic disk.  Thus, Model~3 represents the
opposite extreme of supernova distribution.  
Because this model is perhaps 
most sensitive to overcooling and ``weighting down'' of the 
enriched material due to the relatively high mass 
contained in the supernovae (since few go off inside previously 
evacuated regions), we also test a version of this model (Model 3b) that 
enforces a maximum gas density for all the supernova events 
($\rho_{SN} < 10^{-26}$ g cm$^{-3}$).  
If a supernova region exceeds this density ceiling after applying 
the usual input mechanism, the gas density inside this 
region is uniformly reduced until this criterion is satisfied.  Any 
gas thus removed from the supernova region is redistributed into 
a shell of 2 zones thickness around the supernova so as to 
conserve mass.  This technique is effectively equivalent to requiring 
that all supernovae go off inside evacuated regions and should 
reduce any overcooling or weighting down of the supernova.  
Furthermore, since Model 3 
is probably more sensitive to resolution effects than any other, 
we also use this model 
to test a lower resolution ($128\times128\times64$) simulation (Model 3c).  
Model~4 is an intermediate case between Models 1 and 3 
with the supernovae spread over 30\% of the disk.  We consider a
longer duration (100 Myr) for the supernovae input phase in Models 3 and 4 
to take into account the longer dynamical timescale and 
greater physical distribution of the supernovae sites.  

The supernova rate in Models~1-4 (30 Myr$^{-1}$) is significantly smaller 
than the presently observed
supernova rate in the Milky Way ($10^4$ Myr$^{-1}$).  
However, since the gas disks in these models are comparable in area 
to that of the Milky Way and have average surface density only about a 
factor of 3 smaller, one might expect a more comparable supernova 
rate, perhaps within a factor of 10 \citep{Kennicutt98}.  
Models~5 and 6, therefore, are similar to
Models~1 and 3, except that the energy input rate is increased by an
order of magnitude.  In Model 7 the energy input rate is increased by
yet another order of magnitude.  In Model~8 the mass of the system is
reduced by an order of magnitude; otherwise this model is similar to
Model 3 with an energy input rate equivalent to 30 supernovae per Myr
spread out over 80\% of the disk.  This model allows us to examine the
dependence of the loss of enriched material upon galaxy mass.

\begin{figure}
\epsscale{1.0}
\plottwo{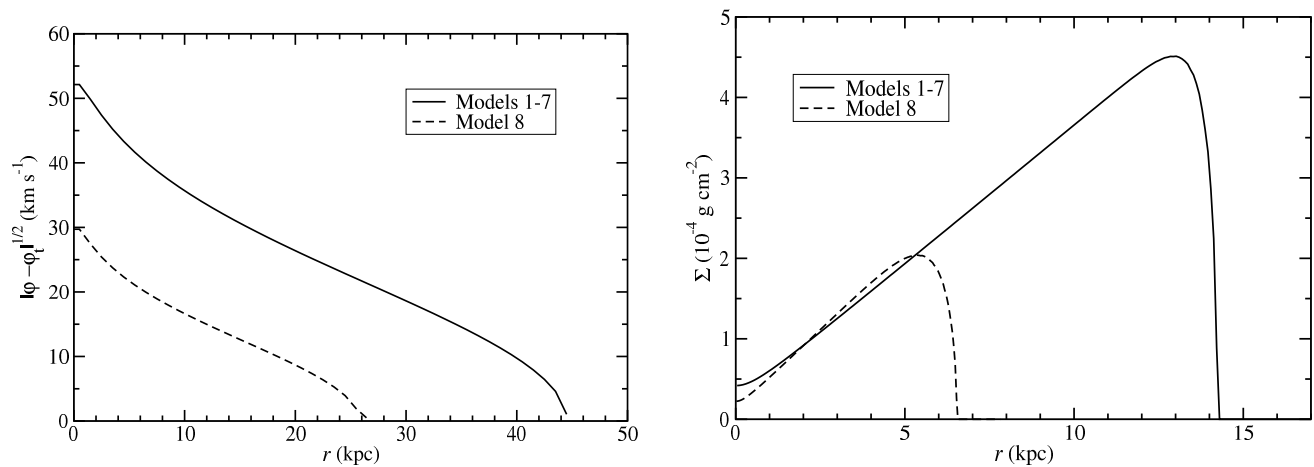}{f3b.eps}
\caption{({\it a}) Square root of the absolute value of the 
normalized gravitational
potential as a function of radius; 
({\it b}) integrated column density as a function of radius in the midplane 
of the disk for the galaxies in 
Models 1-7 ({\it solid curves}) and Model 8 ({\it dashed curves}).
}
\label{fig:galaxies}
\end{figure}

In Figure \ref{fig:galaxies}, we plot the square root of the absolute
value of the normalized gravitational potential ($\vert \phi - \phi_t
\vert^{1/2}$) as a function of radius for the two galaxy masses
considered.  This function gives a good estimate of the local escape
velocity.  At the end of each simulation we use this to estimate the
amounts of unenriched and enriched gas that will ultimately escape the
galaxy, as described in the next section.  We also plot the 
integrated column density ($\Sigma = \int \rho dz$) as a function 
of radius in the midplane of the disk.  This shows that the flaring of 
the disk requires that supernovae going of a larger radii must push through 
a higher column density of material than supernovae occurring near the 
core.  This has a noticeable effect on our results.

\subsection{Loss of Enriched Material}
\label{sec:loss}
We estimate the ejection efficiencies of galactic gas and 
supernova-enriched material by quantifying how much of each material 
is traveling faster than the local escape speed of the dark matter 
halo at the end of the simulation.  
In this way, we define the mass-ejection efficiency as 
\citep{MF99}
\begin{equation}
\xi = (M_{g,esc}+M_{g,lost})/M_{g,i} ~,
\end{equation}
where $M_{g,esc}$ is the gas mass on the grid moving at speeds greater 
than the escape speed, $M_{g,lost}$ is an estimate of the gas mass 
lost from the grid at greater than the escape speed, 
and $M_{g,i}$ is the initial gas mass of the galaxy.  
Similarly, we define the metal-ejection efficiency as
\begin{equation}
\xi_Z = (\mathcal{T}_{esc}+\mathcal{T}_{lost})/\mathcal{T}_{SN} ~,
\end{equation}
where $\mathcal{T}_{SN}$ is the sum of the tracer injected into each
simulation and $\mathcal{T}_{esc}$ and $\mathcal{T}_{lost}$ are
defined analogously to $M_{g,esc}$ and $M_{g,lost}$.  Because our
galaxies are not in initial pressure balance at the disk edge, they
tend to expand sideways at the sound speed during the course of the
simulations.  Furthermore, since the larger galaxy models (Models 1-7)
initially fill up a large fraction of the grid, they tend to lose gas
off of the grid due solely to this sideways expansion.  We do not
include this loss in our estimate of $M_{g,lost}$.  We evaluate $\xi$
and $\xi_Z$ at the end of each run (50, 75, or 100 Myr) 
and report the results in Table \ref{tab:eject}.

\section{Results}
\label{sec:results}

\subsection{Model 1}
\label{sec:model1}

\begin{figure}
\epsscale{1.0}
\plotone{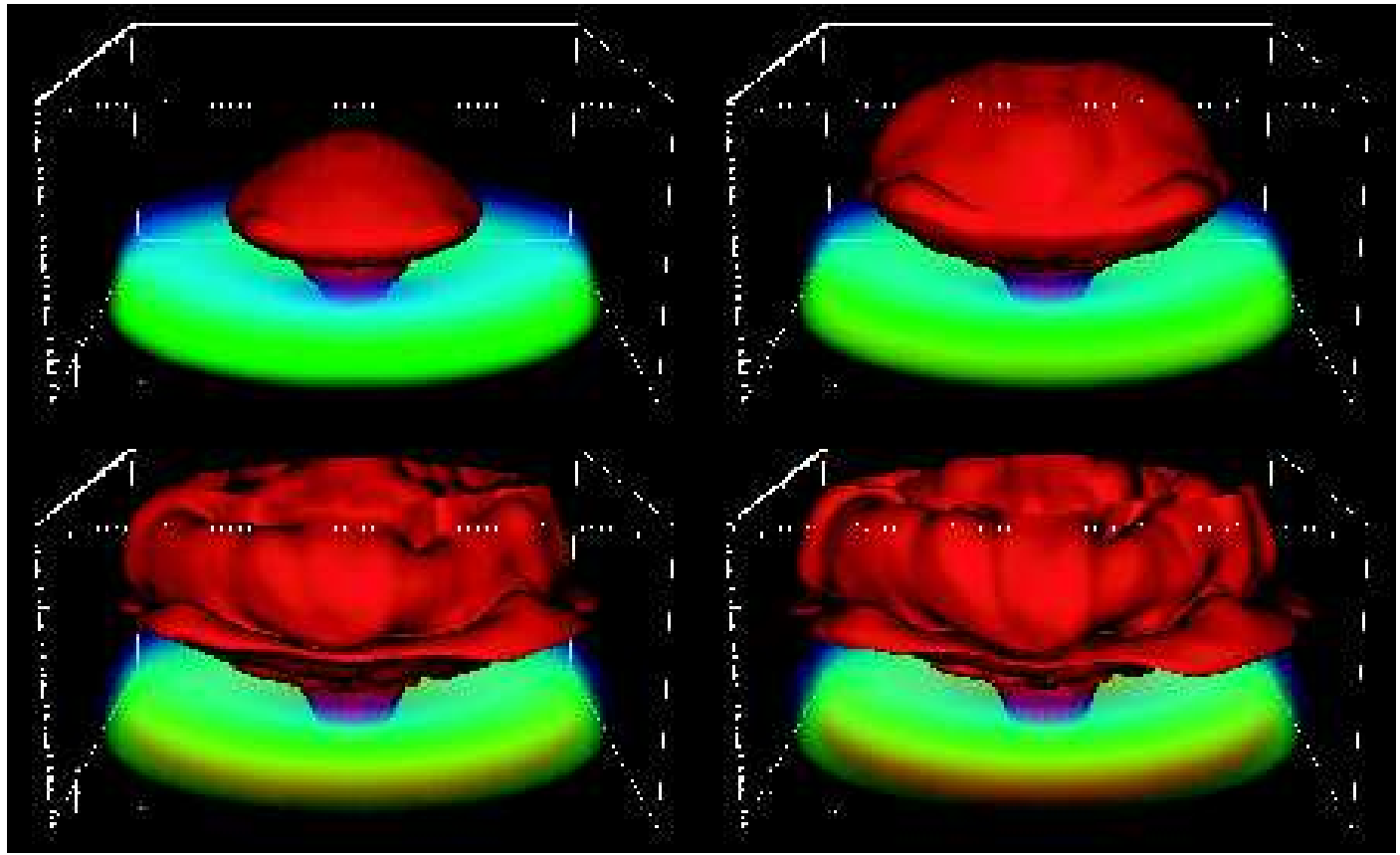}
\caption{Volume rendering of gas density plus an isosurface (in red) of metal 
tracer for Model 1 at $t=25$, 50, 75, and 100 Myr.  
The metal tracer surface is 
set at 0.1\% of the input level.  The box is $30\times30\times15$ kpc$^3$; 
the major tick marks are spaced 4.5 kpc apart 
in the $x$- and $y$-directions and 2.25 kpc apart in the $z$-direction.
}
\label{fig:mod1}
\end{figure}

The density and tracer evolution for Model~1 are shown in
Figure~\ref{fig:mod1}.  In this figure, the density distribution of
ordinary gas is illustrated using volume visualization.  Simultaneously
we plot, in red, an isosurface of the tracer gas set at $10^{-3}$ of
the input level.  Similar figures are presented for each of our
models.

\begin{figure}
\epsscale{0.5}
\plotone{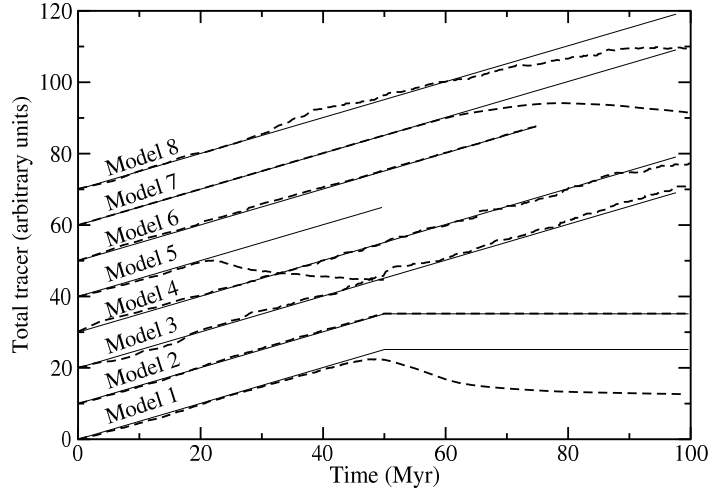}
\caption{Plot of total tracer contained within $R_d$ 
as a function of time for all models.  The data are adjusted to have 
the same slope and then offset to make the plot more legible.  The thin solid 
lines are provided as a guide of the average tracer input rate 
for each model.  Note that, for Models 1 and 2, supernova input ends after 
50 Myr.  The remaining models have continuous supernova input for the 
duration of the simulations.
}
\label{fig:tracer}
\end{figure}

As can be seen in Figure \ref{fig:mod1}, the
supernova energy input in Model 1 
causes a rapid ``chimneying'' up the symmetry axis
of the galaxy.  The preferential motion of enriched gas along the rotation axis
leads to a substantial loss.  The time history of this loss is 
presented in Figure \ref{fig:tracer}, alongside data from our other models.
This figure tracks the cumulative 
amount of tracer material inside of $R_d$ for each model as a function 
of time.  Each model is adjusted to the same slope and the results of the
models are offset from each other for 
clarity.  The thin solid lines show the adjusted average input rate for each 
model.  Although this plot is useful for illustration, it does not 
account for enriched material that is traveling 
faster than the escape speed but still located within $R_d$.  Such 
material is expected to eventually escape the galactic potential.  We 
will return to this point below.

\begin{figure}
\epsscale{0.5}
\plotone{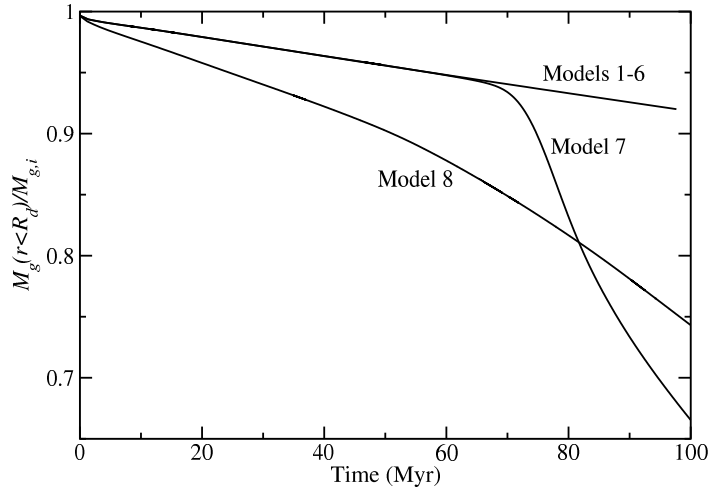}
\caption{Plot of total gas mass ($M_g$) contained within $R_d$ 
as a function of time for all models.  The data are normalized to
the initial gas mass of each galaxy.  Note that some mass escapes 
beyond $R_d$ simply because the galaxies are not in initial pressure 
equilibrium with the background at the cut-off radius.  This accounts 
for virtually all of the loss for Models 1-6 and the early loss in Models 7 
and 8.  Only Models 7 and 
8 show a deviation due to the energy input of supernovae at late time 
($t\gtrsim 50$ Myr).
}
\label{fig:totmas}
\end{figure}

In Figure \ref{fig:totmas} we plot, for each of our models, the time history
of the total gas mass contained within $R_d$,
normalized to the initial mass of each galaxy.  Note that some mass is lost
beyond $R_d$ in all models due simply to the fact that these galaxies are not 
initially in pressure equilibrium at the disk edge.  The initial 
non-equilibrium leads to a fractional mass loss that is identical for
Models~1-6, and which has nothing to do with ejection by supernovae.
Only Models~7 and 8 show significantly higher mass loss 
than the more quiescent 
models.  Comparing with Figure \ref{fig:tracer}, we see that 
enriched material is preferentially lost in nearly all of these models.

While Figures \ref{fig:tracer} and \ref{fig:totmas} give a qualitative 
feel for the ejection history of enriched material and gas mass 
within our models, the ejection efficiencies defined in \S \ref{sec:loss} 
give a much more physically motivated estimate.  The results for each 
model are summarized in Table \ref{tab:eject}.  For Model 1, 
although the ejection of enriched material is highly efficient 
($\xi_Z=0.91$), very little mass is lost from the galaxy as a whole 
($\xi=0.023$).  We find this to be true for most of our models.

We note that our results for Model 1 are similar, although not identical, 
to those of \citet{MF99}.  This is not
unexpected, however, given that the current study uses somewhat lower 
resolution and a different cooling function.  The fundamental conclusion,
however, that highly concentrated supernovae trigger substantial loss of
enriched material, remains unchanged.

\subsection{Model 2}
\label{sec:model2}
\begin{figure}
\epsscale{1.0}
\plotone{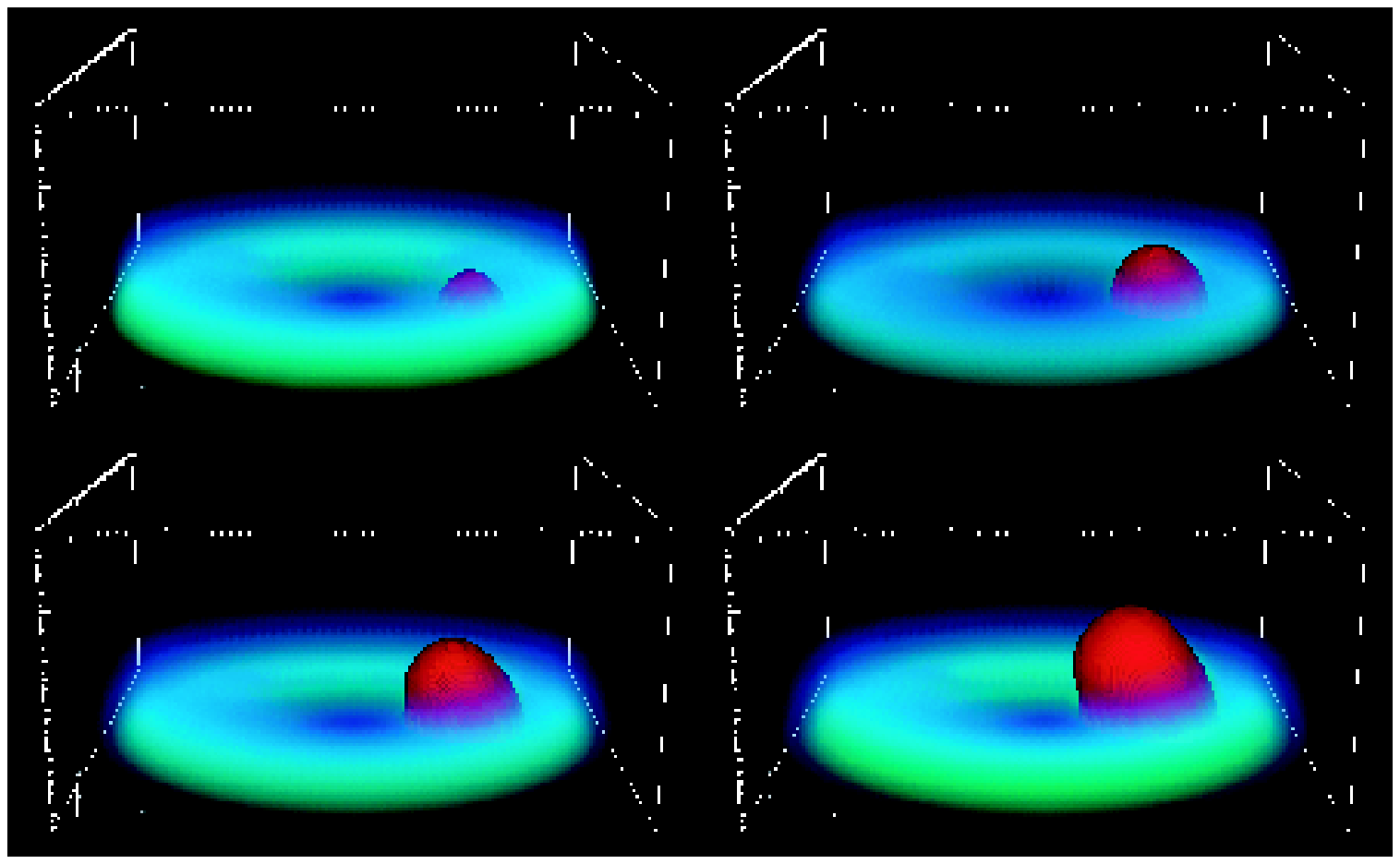}
\caption{Same as Figure \ref{fig:mod1} for Model 2 at 
$t=25$, 50, 75, and 100 Myr.
}
\label{fig:mod2}
\end{figure}

Model 2 is identical to Model 1 except that the supernovae are all
initiated at a fixed point half way out in the disk.  This allows us
to test the dependence of our results on the location of a starburst.
In Figure \ref{fig:mod2} we plot the evolution of the density and
tracer for this model; it is obvious that there are significant
differences from the results of Model 1.  Due to the flaring of the
galaxy, the starburst ejecta in Model 2 must plow through a thicker
layer of gas (with higher column density) before reaching the
IGM.  Thus more supernova energy is expending tunneling to the surface
of the galaxy, leaving less energy to eject the enriched material.
The higher column density also makes cooling more efficient in the 
swept up material, further reducing the energy available to expel 
enriched material.  From Figures \ref{fig:tracer} and
\ref{fig:mod2}, we see that, by the end of the simulation, none of
the enriched material has escaped beyond $R_d$, in stark contrast to
Model 1.  Also, from Table \ref{tab:eject}, we note that, although 
a slightly larger amount of unenriched gas is ejected, much less 
enriched material escapes the galaxy.  This suggests that the location 
of a starburst or supernova can be an important factor in determining 
how the final products of stellar evolution are distributed.

\subsection{Model 3}
\label{sec:model3}

In Model~3 the energy input rate is identical to that of Model~1, but the
supernovae occur over 80\% of the galactic disk.  This distribution 
represents the opposite extremum from Models~1 and 2.  Whereas those models
represent the results of starburst-like activity, Model~3 represents the 
situation of steady, self-regulated star formation throughout the disk.

\begin{figure}
\epsscale{1.0}
\plotone{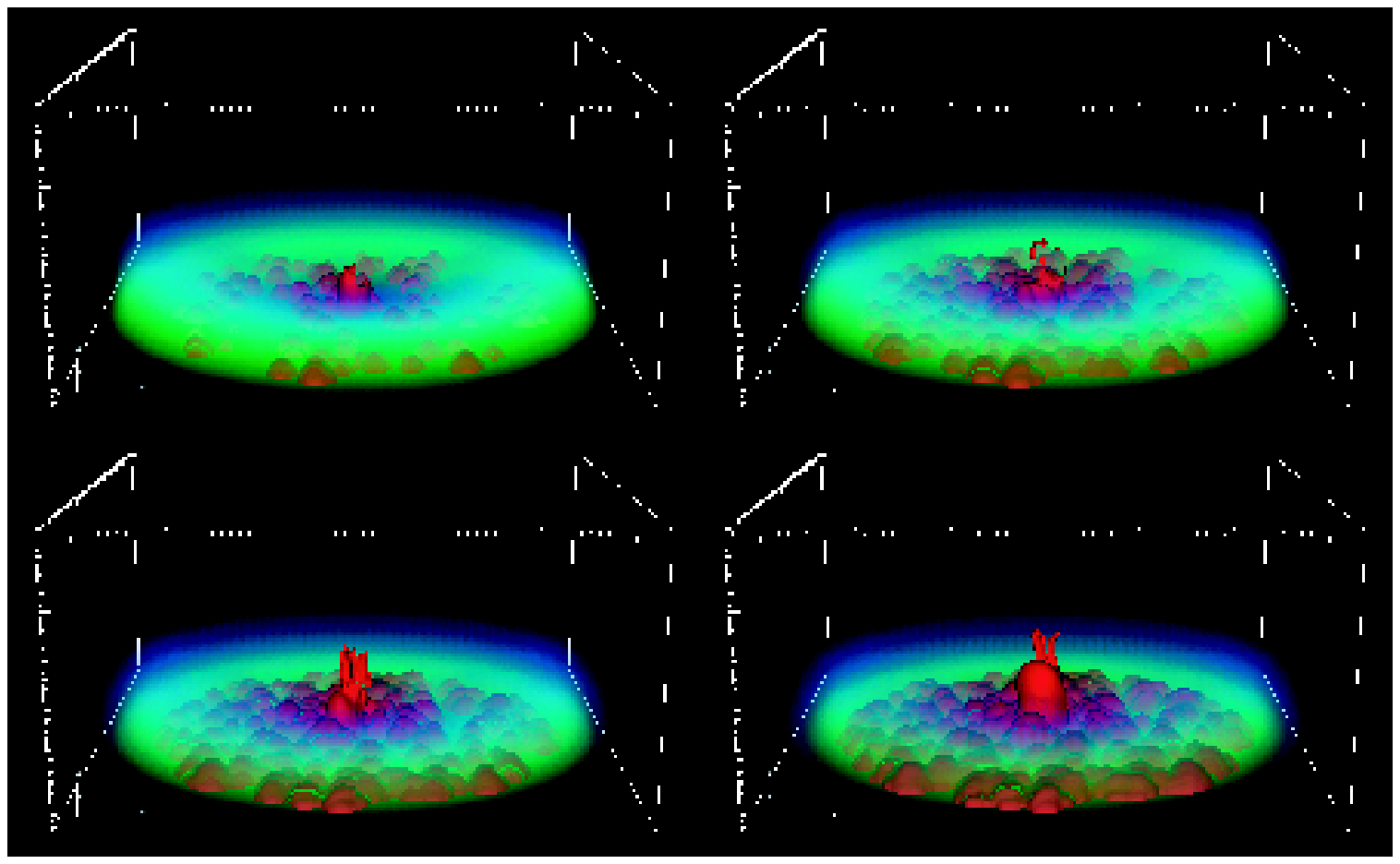}
\caption{Same as Figure \ref{fig:mod1} for Model 3a at 
$t=25$, 50, 75, and 100 Myr.
}
\label{fig:mod3}
\end{figure}

The evolution of the
density and tracer for Model~3a is shown in Figure~\ref{fig:mod3}.
As can be seen in the figure, because the supernovae are spread out over a
large fraction of the disk, they only interact with a few of their closest 
neighbors.  The galactic gas, therefore, has a significant amount
of time to relax between supernova events.  The result is that 
most of the enriched material is retained by the galaxy as indicated
in Figure \ref{fig:tracer}.  This is also supported by the very low 
ejection efficiencies of this model in Table \ref{tab:eject}.  
This model suggests that highly 
distributed star formation will be much less efficient at 
ejecting enriched material than more concentrated starbursts.

Because this model shows the least efficient ejection and is perhaps 
most subject to overcooling and weighting down due to the relatively high mass 
contained in each supernova (since few go off in previously evacuated 
regions), we also test a version of this model that 
enforces an maximum gas density ($\rho_{SN} < 10^{-26}$ g cm$^{-3}$) 
inside the supernova region for all 
events (Model 3b).  
This should tell us if the high mass inside the initial supernova region 
is responsible for the poor ejection efficiency.  
Instead, we find the evolution of this model to be very similar 
to Model 3a.  Although Model 3b does 
show a slightly higher metal ejection efficiency ($\xi_Z=0.24$), the 
difference would not seem to invalidate any of our results.

Model 3c is yet another variant of this basic model.  
Model 3c uses half the normal number of zones 
in each direction; it is intended to test the effect of 
resolution on our conclusions.  
We note that the low resolution model shows less efficient 
metal ejection ($\xi_Z=0.17$) than the higher resolution equivalent.  
The lower ejection efficiency of Model 3c is 
most likely attributable to extra numerical diffusion of the ejecta 
into the surrounding galactic gas.
However, the difference is not dramatic, suggesting that our results are well 
enough resolved to support our general conclusions.  

\subsection{Model 4}

\begin{figure}
\epsscale{1.0}
\plotone{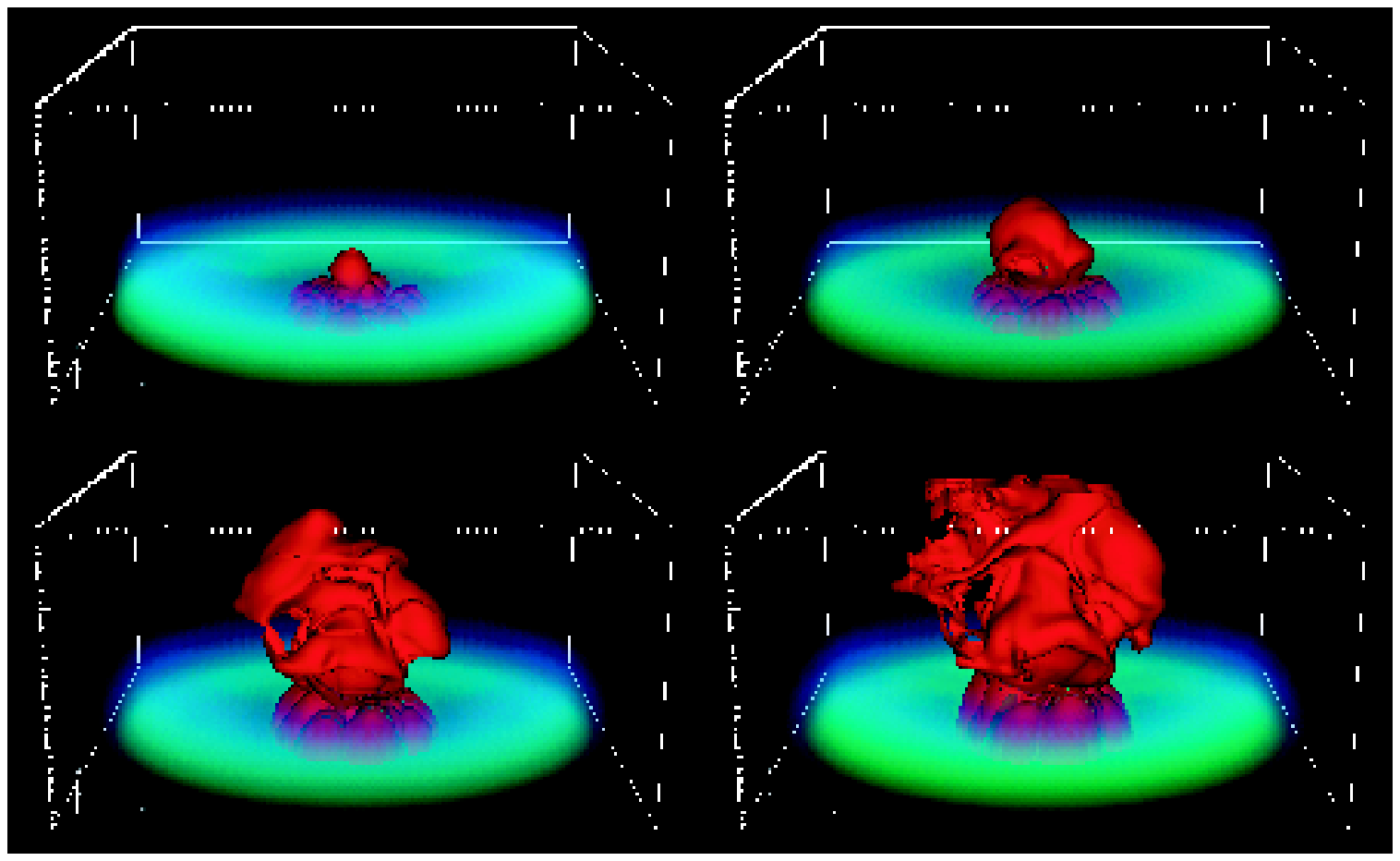}
\caption{Same as Figure \ref{fig:mod1} for Model 4 at 
$t=25$, 50, 75, and 100 Myr.
}
\label{fig:mod4}
\end{figure}

Model~4 is similar to Model~3, except that the supernovae are concentrated 
within the inner 30\% of the disk.  Thus it represents an intermediate 
case between Models~1 and 3.  The evolution of the density and tracer
for this model is shown in Figure~\ref{fig:mod4}.
From Figure \ref{fig:tracer} we see that after 100~Myr, 
only a small amount of tracer has been lost beyond $R_d$, although 
the calculated metal ejection efficiency ($\xi_Z=0.60$) is clearly 
higher than Model 3 but much less than Model~1.  
Thus, even a moderate distribution of supernova energy can 
significantly reduce the ejection efficiency of enriched gas.

\begin{figure}
\epsscale{0.5}
\plotone{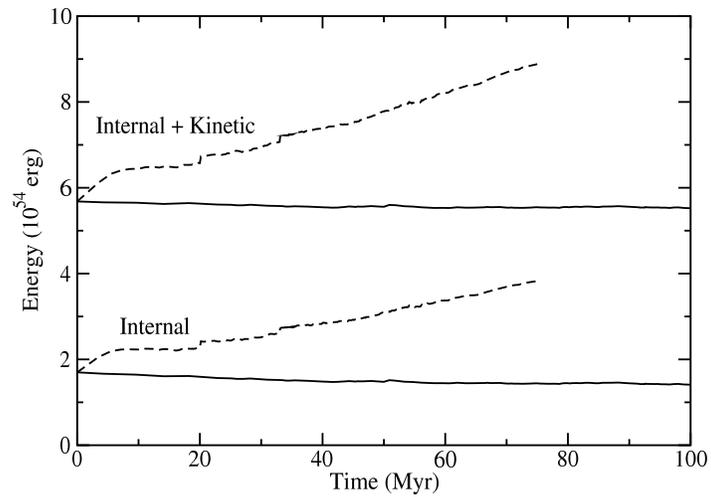}
\caption{Internal and internal+kinetic energy as a function of time 
for Models 3 ({\it solid line}) and 6 ({\it dashed line}).
}
\label{fig:energy}
\end{figure}

Because the total supernova energy input in Models~3 and 4 is less than the
binding energy of the gas, it might be argued that the supernovae have not
had sufficient time to have a significant effect by the end of the
simulations.  In order to examine the possibility that continued supernova
input may change our conclusions, we plot, in Figure~\ref{fig:energy}, the
time evolution of the internal and internal plus kinetic energies of Model~3.
As can be seen, the energy of the model maintains an approximately
steady-state, in which energy is radiated away at the same rate at which it
is input by supernovae.  We obtain very similar results for Model 4.  
These models would not, therefore, evolve significantly if the simulations
were carried out for longer times.

\subsection{Model 5}
\label{sec:Model 5}

\begin{figure}
\epsscale{1.0}
\plotone{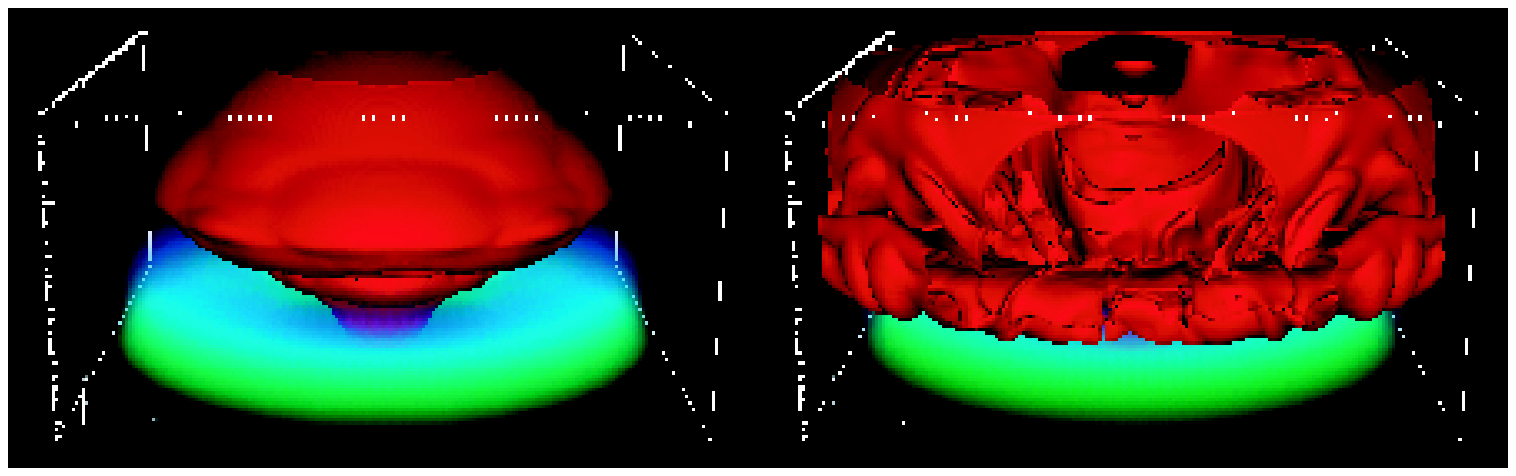}
\caption{Same as Figure \ref{fig:mod1} for Model 5 at 
$t=25$ and 50 Myr.
}
\label{fig:mod5}
\end{figure}

Model~5 is similar to Model 1 except that the energy input rate is
increased by an order of magnitude.  The supernova rate is therefore much
closer to that of the Milky Way than in the previous models, as might be
expected due to the similar surface density of gas and overall disk area
of the models as compared to the Milky Way.  The evolution
of the density and tracer for this model is shown in Figure~\ref{fig:mod5}.  
Similar to Model 1, nearly all of the enriched material chimneys out 
of the galaxy along the symmetry axis, while at the same time, very little
mass is lost from the remainder of the disk.  Thus, our conclusion that 
highly concentrated starbursts result in efficient metal ejection 
without disturbing the bulk of the galaxy does not appear to be 
strongly dependent on the overall energy input rate.

\subsection{Model 6}
\label{sec:Model 6}

\begin{figure}
\epsscale{1.0}
\plotone{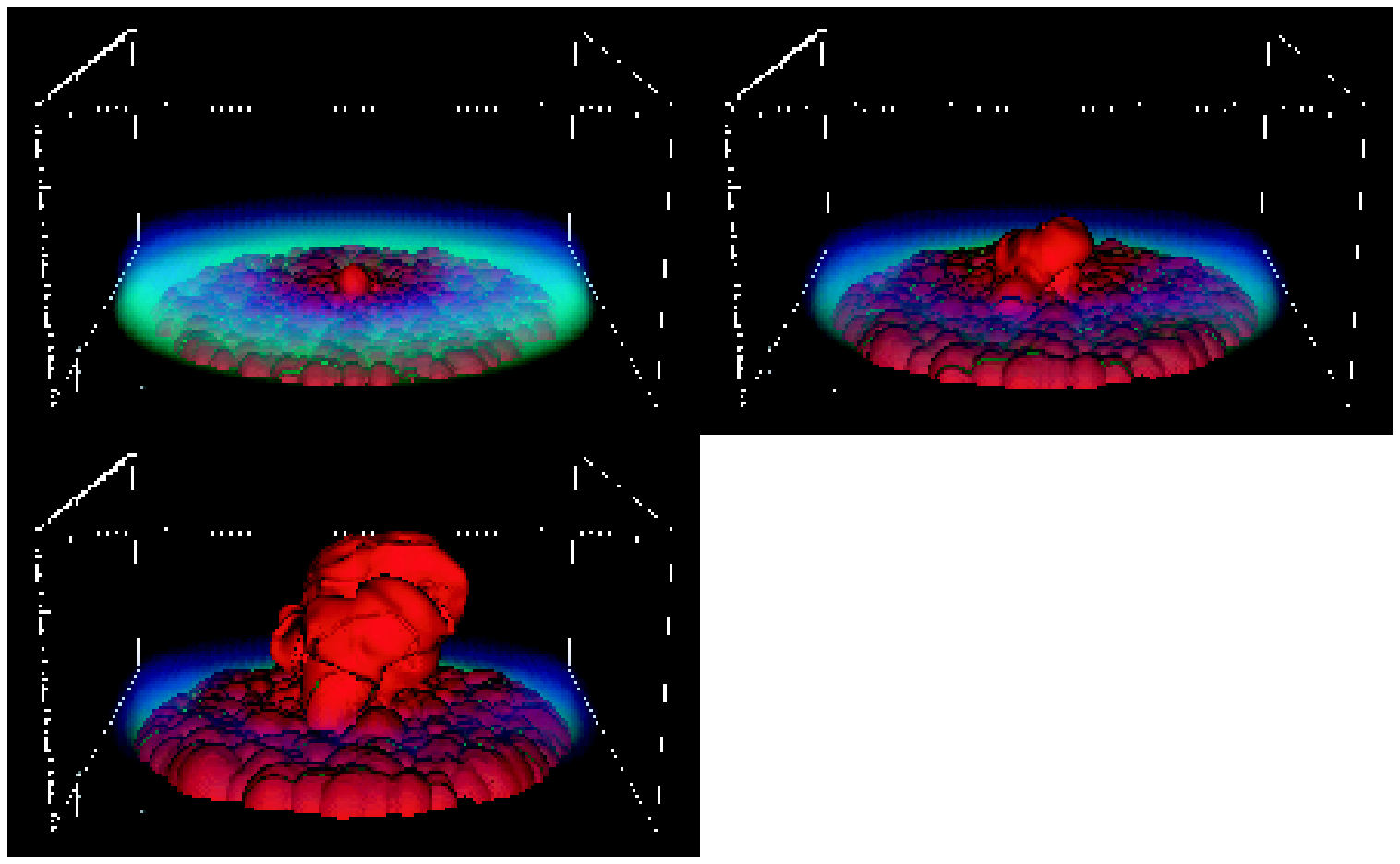}
\caption{Same as Figure \ref{fig:mod1} for Model 6 at 
$t=25$, 50, and 75 Myr.
}
\label{fig:mod6}
\end{figure}

Model 6 has the same supernova rate as Model 5, yet 
the supernovae are spread over 80\% of the disk, as in Model~3. The evolution
of the density and tracer for this model is shown in Figure~\ref{fig:mod6}.
In contrast with Model~3 (Figure~\ref{fig:mod3}), we see that 
the rate of energy input is sufficient to lead to some
``tunneling,'' providing low densities avenues for the loss of some enriched
gas, particularly at small radii.  Nevertheless, 
the loss of tracer for Model~6 is still small ($\xi_Z=0.60$) compared to 
highly concentrated runs and there is very little 
supernova-driven mass loss by the end of the simulation ($\xi=0.096$).  
However, we clearly see in Figure \ref{fig:energy} that the energy 
input has not reached a steady state for this model.  
The energy input rate remains greater than the rate of loss due to 
radiative emission.  
If supernovae were to continue at the modeled rate indefinitely, 
the disk would eventually be destroyed, similar to what we observe 
in Model~7, discussed below.  However, it is more realistic to 
assume that the supernova rate is 
correlated with the gas density in the disk, and thus the supernovae 
would likely shut off before that could happen, in which case the 
disk may remain intact.

\subsection{Model 7}
\label{sec:Model 7}

\begin{figure}
\epsscale{1.0}
\plotone{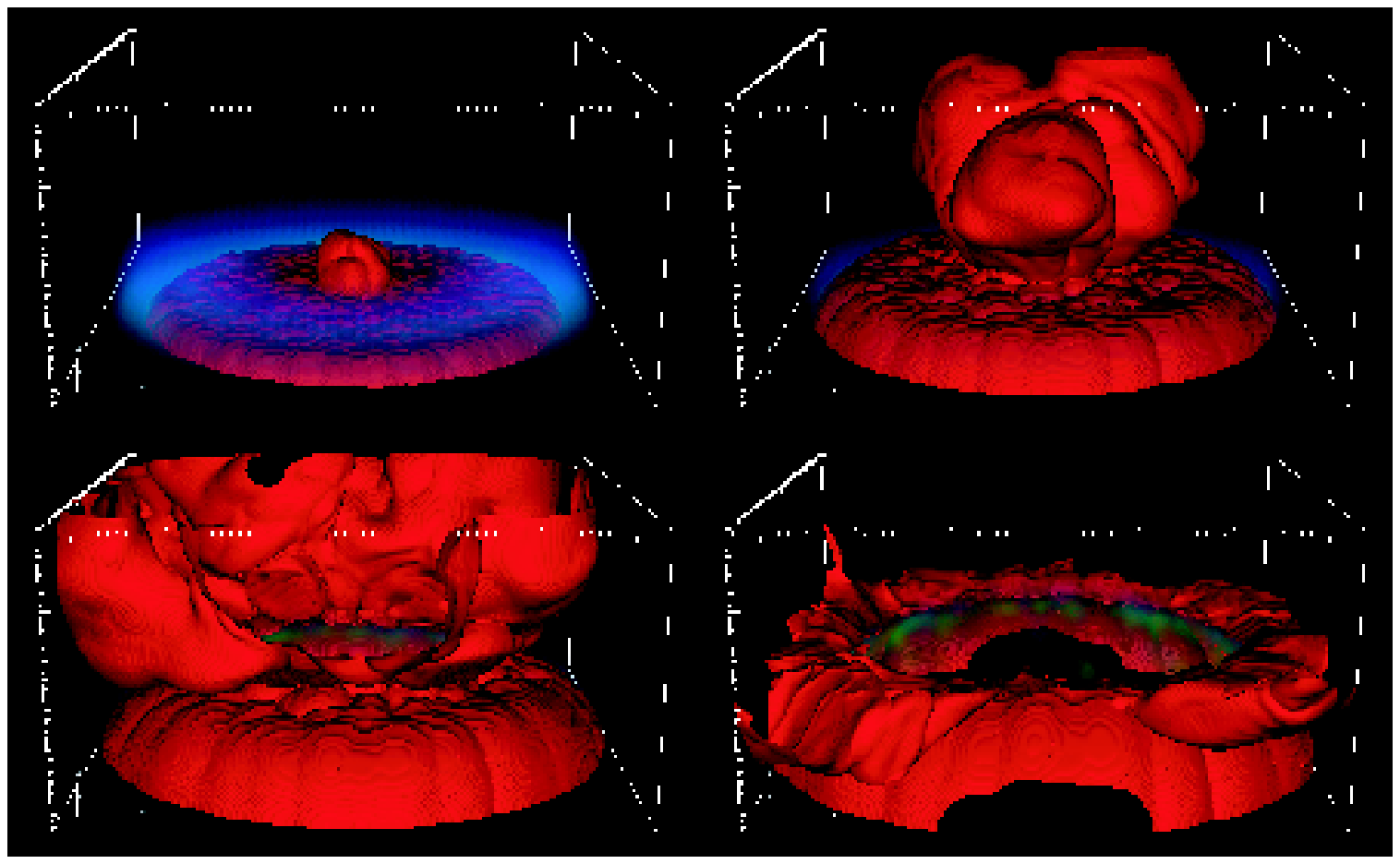}
\caption{Same as Figure \ref{fig:mod1} for Model 7 at 
$t=25$, 50, 75, and 100 Myr.
}
\label{fig:mod7}
\end{figure}

Model~7 is similar to Model 6 except that the energy input rate is
increased by yet another order of magnitude.  The evolution of the
density and tracer for this model is shown in Figure~\ref{fig:mod7}.
The supernova energy input rate of Model~7 greatly exceeds the
radiative efficiency of the galaxy.  The result, as can be seen in
Figures~\ref{fig:tracer}, \ref{fig:totmas}, and \ref{fig:mod7}, is
that after 70~Myr, nearly all of the tracer added to the galaxy so far
is lost, and the disk itself is completely disrupted, leading to
extensive mass loss.  Since we have fixed the duration of the supernovae
injection for this model at 100~Myr, the continuous infusion of 
enriched material ensures a residual
of tracer elements are found even at late epochs (see Fig.~\ref{fig:tracer}).
The results of this model suggest that, for supernovae distributed over a large
fraction of the galaxy, a high efficiency of metal loss will be
accompanied by significant overall mass loss.

\subsection{Model 8}
\label{sec:Model 8}

Model~8 is similar to Model~3 except that the mass of the galaxy is 
an order of magnitude smaller.  This model allows us to explore the dependence
of our results on galaxy mass.  A greater loss of enriched gas relative to
Model~3 is expected for two reasons.  Firstly, the potential of Model~8
is significantly shallower than that of Model~3 (Figure~\ref{fig:galaxies}).
Secondly, while the two models have same supernova energy input rate, the
total radiative emission rate of Model~8 is reduced by more than an order of
magnitude relative to that of Model~3 (due to the lower overall gas 
mass and lower gas density).

\begin{figure}
\epsscale{1.0}
\plotone{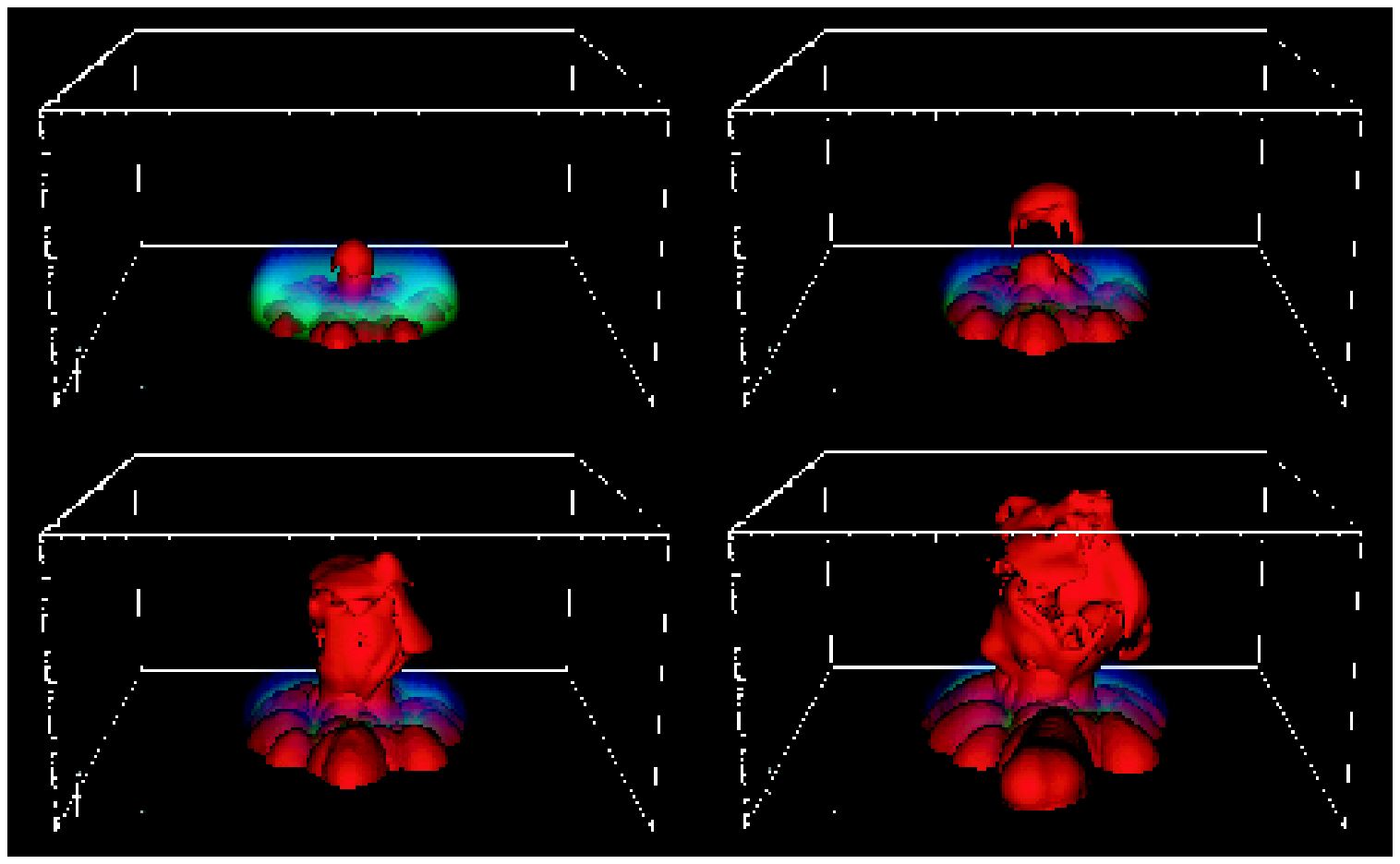}
\caption{Same as Figure \ref{fig:mod1} for Model 8 at 
$t=25$, 50, 75, and 100 Myr.
}
\label{fig:mod8}
\end{figure}

The evolution of the density and tracer for Model~8 is shown in
Figure~\ref{fig:mod8}.  As can be seen, the system is unable to relax
significantly between supernova events.  The gas distribution
therefore becomes highly disturbed, even more so than that of Model~6,
despite the fact that these two models have the same energy input rate
per unit gas mass.  The difference is again due both to the shallower
gravitational potential of Model~8 and its reduced radiative 
efficiency.

The evolution of the tracer for Model~8 shows three distinct
behaviors.  Initially, no tracer is lost to the system.  After
approximately 60~Myr, tracer begins to be lost at an approximately
constant rate.  Late in the simulation, the tracer abundance flattens,
i.e. enriched material is lost to the system at roughly the same rate
it is being added.  As can be seen in Figure \ref{fig:mod8}, the early
losses of tracer are the result of material ``chimneying'' near the
symmetry axis.  The increase in the loss rate at approximately 85~Myr,
occurs when enriched material also begins to expand beyond $R_d$ in
the equatorial plane of the disk (see the last frame of Figure
\ref{fig:mod8}).

\section{Discussion and Implications}
\label{sec:discussion}

We have performed three-dimensional simulations of the evolution of
enriched gas within dwarf disk galaxies.  In these simulations,
supernovae occur at random intervals and at random locations over
prescribed fractions of the disk area.  Our primary interest is to
consider the efficiency of metal ejection for these systems, and its
consequences for the evolution of dwarf galaxies, as we discuss below.


In general, we find that supernovae are less effective at ejecting
enriched material from dwarf disk galaxies than has been suggested in
previous work.  The key difference between the current and the earlier
models \citep{MF99} is that the earlier studies triggered supernovae
either within the cores of the galaxies, or elsewhere within the
systems but still with very high local efficiency.  In those models,
representative of starbursts, the ambient gas was unable to relax
between supernova events, and enriched gas could rapidly ``chimney''
out of the systems.  This same behavior is seen in our starburst
models (1, 2, and 5).  In our remaining models, with supernovae 
distributed over 30 or 80\% of the disk, a substantial fraction of 
the metals are retained.  An important implication is that 
closed-box models for the chemical evolution
of these galaxies may be more appropriate when supernovae are expected 
to be distributed throughout the disk than when they are concentrated in the 
central core or bulge region.  Nevertheless, we find in all our models 
that a galactic outflow is set up in which at least some fraction of 
the available metals is ejected.

\subsection{Comparison with Theoretical Supernova Rates}

In the self-regulated star-formation model of \cite{LM92}, massive
stars ionize and heat surrounding gas, preventing further star
formation until the massive stars evolve off of the main sequence.
This picture leads to a straightforward prediction for the formation
rate of massive stars and the resulting supernova rate.  From
\cite{LM92}, the number of massive stars within a disk out to radius
$R$ is given approximately as
\begin{equation}
N_\ast\sim{\frac{2\pi R^2H}{\frac{4}{3}\pi R_S^3}},
\label{eq:nstars}
\end{equation}
where 
\begin{equation}
H=2^{1/2}c_s/\Omega
\end{equation}
is the scale height of the gas, $\Omega$ is the rotational frequency at
$R$,
\begin{equation}
R_S^3={\frac{Q_\ast}{\frac{4}{3}\pi n^2\alpha_B}}
\end{equation}
is the Str\"omgren sphere radius, $Q_\ast$ is the rate of emission of
photoionizing photons by a massive star averaged over the population
of massive stars, $n$ is the average number density of the gas, and
$\alpha_B$ is the case B recombination coefficient \citep{Ost89}.
Using reasonable values for the above parameters, it is found that
\begin{equation}
N_\ast\sim2000{~\rm R}_{10}^{-7/2}{~\rm M}_{10}^{5/2}{~\rm b}_{.1}^2,
\end{equation}
where R$_{10}$ is the radius of the disk in units of 10~kpc, M$_{10}$
is the total mass of the system in units of 10$^{10}$~M$_\odot$, and
b$_{.1}$ is the gas-to-total mass ratio, in units of 0.1.  An L$_\ast$
galaxy, such as the Milky Way today, is predicted to have supernova
rates on the order of one per century, close to the observed rate.

For galaxies such as the one used for Models~1-7, supernova rates
of $\sim$ 100~Myr$^{-1}$ would be expected, intermediate to the rates
applied in Models~3 and 6.  It might be expected, therefore, that the
loss of enriched material for a galaxy of that mass undergoing
self-regulated star formation would lie between 21\% and 60\%.  
We note, however, that the supernova rate would not remain at this value 
throughout the history
of the galaxy.  Rather, it would decrease as the
square of the gas mass, assuming that the gas scale height remained
constant with time.  The rate of enrichment would, therefore, decrease
linearly with the gas fraction, as gas is converted into stars.

A galaxy such as that in Model~8 would be predicted to have a
supernova rate $\sim 6$~Myr$^{-1}$, several times less than applied in
that model.  Therefore, the loss of enriched material for such a
galaxy undergoing self-regulated star formation may be significantly
less than seen for Model~8 (53\%).

\subsection{Implications for Dwarf Galaxy Evolution}

In the least massive galaxies (total masses $\lesssim10^9$~M$_\odot$),
simulations of both disk and spheroidal systems find that enriched gas
from supernovae may be largely lost to the systems \citep{MF99, SGP03,
FMAL03}.  Such evolution is supported by observations which find poor
self-enrichment efficiency in many low-mass systems
\citep{Mateo98,Grebel01,H01,PGWCD03}.

In more massive systems, however, our current simulations find that
the supernova rates expected for self-regulated star formation do not
lead to significant loss of enriched gas when the supernovae are
spread over a reasonable fraction of the disk.  Such galaxies would,
therefore, be expected to be efficiently self-enriched.  Highly
efficient loss of enriched gas occurs only if star formation occurs in
concentrated bursts or significantly exceed estimated self-regulated
rates.  In those cases, the inability of the gas to relax between
supernova events allows substantial loss.

Our conclusions apply primarily to the outputs of Type II supernovae,
the progenitors of which are massive, young stars.  The short
lifetimes of these objects do not allow for sufficient time to migrate
large distances from their places of birth.  Therefore, they are
generally associated with regions of active star formation.  In disk
galaxies, these are near the midplanes, where the gas density is
highest.  The placement of the supernovae in the midplanes of our
models is, therefore, a good representation to the actual distribution
of Type II supernovae within galaxies.  The progenitors of Type I
supernovae, on the other hand, are older systems.  Type I events may
therefore occur further away from the disk midplane, giving them a
better chance to eject enriched material from the galaxy.  This is
especially interesting, given that Type I supernovae are believed to
be the dominant source of iron in galaxies, accounting for
approximately 70\% of the total production \citep{Nometal84,
Wooetal86, WW86, Matteucci88}.  It is possible that although dwarf
galaxies are able to retain the heavy elements produced by Type II
events, they may still be depleted in iron.  
This picture is supported by two-dimensional numerical simulations 
of Type I and Type II supernovae \citep{RMD01}.  

\begin{acknowledgements}
This work was performed under the auspices of the U.S. Department of
Energy by University of California, Lawrence Livermore National
Laboratory under Contract W-7405-Eng-48.  This work is partially
supported by NASA through an astrophysical theory grant NAG5-12151.
\end{acknowledgements}


\clearpage
\begin{deluxetable}{ccccccc}
\tablewidth{0pt}
\tablecaption{Model Galaxy Parameters \label{tab:models}}
\tablehead{
\colhead{Model} &
\colhead{$M_g$} &
\colhead{Energy input rate} &
\colhead{Energy/event} &
\colhead{$R_{SN}/R_d$} &
\colhead{SN input phase} &
\colhead{Simulation stop} \\
\colhead{} &
\colhead{(M$_\odot$)} &
\colhead{(10$^{51}$ erg Myr$^{-1}$)} &
\colhead{(10$^{51}$ erg)} &
\colhead{} &
\colhead{(Myr)} &
\colhead{(Myr)}
}
\startdata
1  & 10$^9$ & 30   & 1  & 0.0 & 50 & 100 \\ 
2  & 10$^9$ & 30   & 1  & 0.0\tablenotemark{a} & 50 & 100 \\ 
3  & 10$^9$ & 30   & 10 & 0.8 & 100 & 100 \\ 
4  & 10$^9$ & 30   & 10 & 0.3 & 100 & 100 \\ 
5  & 10$^9$ & 300  & 10 & 0.0 & 50 & 50 \\ 
6  & 10$^9$ & 300  & 10 & 0.8 & 75 & 75 \\ 
7  & 10$^9$ & 3000 & 1  & 0.8 & 100 & 100 \\ 
8  & 10$^8$ & 30   & 10 & 0.8 & 100 & 100 
\enddata
\tablenotetext{a}{Offset from center by $R_d/2$.}
\end{deluxetable}

\begin{deluxetable}{ccc}
\tablewidth{0pt}
\tablecaption{Gas and Metal Ejection Efficiencies \label{tab:eject}}
\tablehead{
\colhead{Model} &
\colhead{$\xi$} &
\colhead{$\xi_Z$}
}
\startdata
1  & 0.023 & 0.91 \\ 
2  & 0.032 & 0.47 \\ 
3a & 0.027 & 0.21 \\ 
3b\tablenotemark{a} & 0.027 & 0.24 \\ 
3c\tablenotemark{b} & 0.027 & 0.17 \\ 
4  & 0.028 & 0.60 \\ 
5  & 0.016 & 0.99 \\ 
6  & 0.096 & 0.60 \\ 
7  & 0.98  & 0.99 \\ 
8  & 0.064 & 0.53    
\enddata
\tablenotetext{a}{Density ceiling enforced on all 
supernovae.}
\tablenotetext{b}{$128\times128\times64$ resolution.}
\end{deluxetable}


\clearpage

\end{document}